\documentclass[final]{elsarticle}
\usepackage{hyperref,amsmath}
\usepackage{graphicx}

\journal{Granular Matter}

\begin{document}

\begin{frontmatter}

\title{Hourglass of constant weight as an illustrative example of a system with a variable mass}

\author{Vladimir V. Kassandrov}
\ead{vkassan@sci.pfu.edu.ru} 
\address{Institute of Gravitation and Cosmology, Peoples' Friendship University of Russia, 117198 Moscow, Russia}

\author{Zurab K. Silagadze}
\ead{Z.K.Silagadze@inp.nsk.su}
\address{Budker Institute of Nuclear Physics, 630 090, Novosibirsk, Russia } 
\address{Novosibirsk State University, 630 090, Novosibirsk, Russia}

\begin{abstract}
We provide an alternative approach to a problem posed in the article ``Hourglass of Constant Weight" by Volker Becker and Thorsten P\"{o}schel, which appeared in {\it Granular Matter} {\bf 10} (2008), 231–232, i.e. 14 years ago. Our goal to return to this article after so much time is purely pedagogical. Although there is a slight and subtle inaccuracy in the derivation of equation (8) in the work of Becker and P\"{o}schel, which we correct, we use this circumstance only to give a detailed pedagogical exposition of an excellent physical problem and to show the pedagogical usefulness of the concept of momentum flux when considering the mechanics of systems with variable mass.
\end{abstract}

\end{frontmatter}

\section{Introduction}
Does a running hourglass change its weight? It may look like a question in a science quiz, but in fact this problem has a much broader pedagogical value. It illustrates the intricacies of applying physical laws to explain real phenomena, as students' intuition often goes astray, failing to take into account some important aspects of the problem, and the first solutions that come to mind are wrong. 

We use a variant of the hourglass problem to illustrate the concept of momentum flux in the dynamics of systems with variable mass. Despite the practical importance of systems with variable mass, textbooks give this section of mechanics less attention than it deserves, often limiting themselves to considering only the rocket problem. We hope that considering the hourglass problem as a  problem of variable mass systems will be instructive and useful for students.

The structure of the manuscript is as follows. In the next section, we give a general, albeit rather detailed, overview of the classic hourglass problem and proposed solutions. Then we will focus on  the modification of the problem described in Refs.~\cite{Becker2008} and \cite{Tuinstra_2010}. The last reference uses a setup of a hourglass with 230 holes to increase the rate of the sand flow in it, and therefore the magnitude of increase in the weight of the hourglass, allowing it to be accurately measured. For definiteness, and having in mind a possible experimental realization, we use the hourglass parameters described in Ref.~\cite{Tuinstra_2010}, although strictly speaking, for the theoretical reasoning we will focus on, this is not necessary and a 230-hole setup might look unnecessarily complicated. In any case, in the third section, we show that even for a 230-hole setup, the fraction of sand in free-fall is negligibly small.

In the fourth section, we re-obtain the result of Ref.~\cite{Becker2008} for the weight of an hourglass consisting of two compartments connected by springs with elastic constants $k/2$ each. Then, in the next section we obtain the equation of motion of the lower compartment, considering it as a variable mass system with non-zero momentum flux. The calculation of momentum flux becomes more tricky if we change the boundary of the variable mass system, as described in the sixth section. However, this will allow us to reveal a subtle inaccuracy in the derivation of the equation of motion in Ref.~\cite{Becker2008}, which we will correct in the next section.

In the last sections, we provide some general concluding remarks and cannot resist the temptation to demonstrate the power of the momentum flux concept by giving a simple derivation of the equation of motion in another fascinating variable-mass problem, the motion of a leaky tank car.

\section{The hourglass problem}
Whether or not an hourglass changes its weight, measured by sensitive scales, while running is a fascinating pedagogical problem \cite{Shen_1985,Tuinstra_2010,Redmount_1998,Becker2008,Sack_2017}. This challenge is often used at various physics olympiads and pre-olympiad competitions \cite{Ong_1990,Saurabh_2020,Kalda_2018}. Surprisingly, the "canonical" solution given in Refs.~\cite{,Reid_1967,Ong_1990,Saurabh_2020}, as well as in well-known popular books \cite{Walker_1977,Perelman_1992} are wrong (in the second edition of Ref.~\cite{Walker_1977} (namely, Ref.~\cite{Walker_2006}) the solution is corrected). The popular book Ref.~\cite{Gardner_1989} also incorrectly states, with reference to Ref.~\cite{Reid_1967}, that the weight of an running hourglass remains the same, as if the sand were not pouring.

The canonical solution goes as follows. Let at a given moment of time the free-falling time of the sand is $\tau$. Assuming constant rate of the flow $\mu$, this means that $\mu\tau$ mass of the sand is in a free-fall and thus the corresponding reading of the scales should be reduced by $\mu\tau g$ (note that the calculations and reasoning below are not applicable at the very beginning and at the very end of the granular flow, where $\mu$ is not constant). However, during free fall, a piece of sand gains momentum, which is transferred to the bottom part of the hourglass when that piece of sand stops. Let us estimate the corresponding force due to this momentum transfer. The amount of sand $\mu dt$ that passes through the sieve in an infinitesimal time $dt$, before the impact has a velocity $V=g\tau$ and hence a momentum $dp=\mu dt V$. The length of this column of sand before impact is $dl=\frac{1}{2}g[\tau^2-(\tau -dt)^2]\approx g\tau dt$ and hence it stops during the time $dt^\prime=dl/V= dt$. Therefore, the force acting on the lower part of the hourglass due to the corresponding transfer of momentum is equal to $\frac{dp}{dt^\prime}=\mu V=\mu\tau g$ and fully compensates for the weight loss of the hourglass indicated above. So the hourglass should weigh the same whether the sand is falling or not.

\begin{figure}
    \centering
    \includegraphics[scale=0.6]{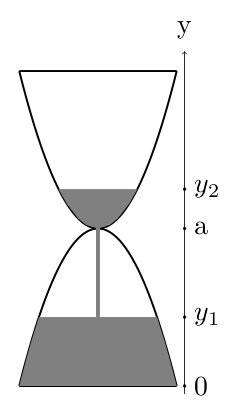}
    \caption{Schematic diagram explaining the meaning of the coordinates $y_1$, $y_2$ and $a$.}
    \label{fig0}
\end{figure}

No matter how clever it looks, this canonical solution is incorrect, neglecting some circumstances that are physically significant for this problem \cite{Shen_1985}. The first circumstance is that the sand leaves the top part of the hourglass and enters into free fall with nonzero speed $V_0$. Correspondingly, $V=V_0+g\tau$ and $dl=V_0\tau+\frac{1}{2}g\tau^2-[V_0(\tau-dt)+\frac{1}{2}g(\tau -dt)^2]\approx (V_0+g\tau) dt=Vdt$. The second circumstance is that the sand moves both in the upper and lower compartments of the hourglass. If the sand level in the bottom part moves upwards with a speed of $V_1$, then the stopping time is modified as follows
\begin{equation}
dt^\prime=\frac{dl}{V+V_1}=\frac{V}{V+V_1}\,dt.
\label{eq_dt}
\end{equation}
Then
\begin{equation}
\frac{dp}{dt^\prime}=\mu V\frac{dt}{dt^\prime}=\mu\left(V+V_1\right)=\mu\left(V_0+g\tau+V_1\right).
\label{eq_dF}
\end{equation}
As we see, the cancellation is not complete and $\Delta F_1=\mu \left( V_0+V_1\right )$ part of the  downward force will remain uncompensated. However, this is not the complete story. There is a third circumstance: the sand in the top part of the hourglass accelerates towards the orifice. The net effect of this acceleration is that some of the sand $\mu dt$ is transferred from the top level, where it has a velocity of $V_2$, to the hole, where it has a velocity of $V_0$. The force causing the corresponding change in momentum $\Delta p=\mu dt\left (V_0-V_2\right )$, according to Newton's third law, entails an upward force $\Delta F_2=\mu \left( V_0-V_2\right )$ on the hourglass (gravity causes this change in momentum of the small piece of sand $\mu dt$ only indirectly through the remaining bulk of the sand in the upper compartment, the direct effect of gravity $\mu dt g$ on this small piece of sand is infinitesimal). Combining $\Delta F_1$ and $\Delta F_2$, we see that the scales will show that the running hourglass is heavier than the stationary hourglass by amount
\begin{equation}
\Delta F= \Delta F_1 - \Delta F_2=\mu \left(V_1+V_2\right ).
\label{Delta_F}
\end{equation}
The mass of sand in the hourglass is
\begin{equation}
m=\rho\int\limits_0^{y_1} A(y)dy+\rho\int\limits_a^{y_2} A(y)dy+\int\limits_{y_1}^a \tilde\rho(y)\tilde A(y)dy,
\label{m_sand}
\end{equation}
where $A(y)$ is the cross-sectional area of the hourglass at height $y$, $\rho$ is the density of the sand in the upper and lower parts of the hourglass, $y_1$ is the height of the sand level in the lower compartment, $a$ is the height of the hole, $y_2$ is the height of the sand level in the upper compartment (see Fig.\ref{fig0}), $\tilde A(y)$ is the cross-sectional area of the falling sand column at height $y$, and $\tilde\rho(y)$ is sand's variable density in it. 
Since $m$ is a constant, differentiating (\ref{m_sand}) we get
\begin{equation}
\left [ \rho A(y_1)-\tilde\rho(y_1)\tilde A(y_1) \right ] \dot{y}_1+\rho A(y_2)\dot{y}_2=0.
\label{mu_balance}
\end{equation}
This relation can be obtained in another way too. Since the portion of sand with mass $\mu dt$ stops in time $dt^\prime$, and taking into account (\ref{eq_dt}) and $V_1=\dot{y}_1$, we can write $\mu dt=\rho A(y_1)\dot{y}_1dt^\prime$, or
\begin{equation}
\mu=\rho A(y_1)\dot{y}_1\,\frac{V}{V+\dot{y}_1}.    
\label{temp1}
\end{equation}
But $\mu=\tilde\rho(y_1)\tilde A(y_1)V$. Expressing $V$ from this relation and substituting into (\ref{temp1}) results in
\begin{equation}
\mu+\tilde\rho(y_1)\tilde A(y_1)\dot{y}_1=\rho A(y_1)\dot{y}_1    
\end{equation}
which is identical to (\ref{mu_balance}), since obviously $\mu=\rho A(y_2) V_2=-\rho A(y_2)\dot{y}_2$.

We expect $\tilde \rho(y_1)<\rho$ and $\tilde A(y_1)\ll A(y_1)$. Therefore, the second term in (\ref{mu_balance}) can be neglected and the relation takes the form
\begin{equation}
\rho A(y_1)\dot{y_1}=-\rho A(y_2)\dot{y_2}=\mu,
\label{eq_mu}
\end{equation}
which implies
\begin{equation}
V_1=\dot{y}_1=\frac{\mu}{\rho A(y_1)},\;\;\; V_2=-\dot{y}_2=\frac{\mu}{\rho A(y_2)},   
\end{equation}
and Eq.(\ref{Delta_F}) becomes
\begin{equation}
\Delta F=\frac{\mu^2}{\rho}\left ( \frac{1}{A(y_1)}+\frac{1}{A(y_2)} \right ).    
\end{equation}
This relation was first obtained in Ref.~\cite{Shen_1985} by evaluating the acceleration of the center of mass of an hourglass. Since then, this method has become the standard for dealing with the hourglass problem \cite{Tuinstra_2010,Redmount_1998,Becker2008,Sack_2017}. Namely, the position of the center of mass $y_{C.M.}$ is given by the equation
\begin{equation} 
My_{C.M.}=\int\limits_0^{y_1}\rho y A(y)dy+\int\limits_a^{y_2}\rho y A(y)dy+
\int\limits_{y_1}^a\tilde\rho(y) y \tilde A(y)dy+C,
\label{eq_CM}
\end{equation}
where $M$ is the total mass of the hourglass and $C$ is a constant corresponding to the contribution of the hourglass framework. Differentiating (\ref{eq_CM}) and using (\ref{mu_balance}) along with $\rho A(y_2)\dot{y}_2=-\mu$, we get
\begin{equation}
MV_{C.M.}=-\mu (y_2-y_1).   
\end{equation}
Therefore, the center of mass velocity is negative, as it should be. However, the center of mass accelerator is positive:
\begin{equation} 
\Delta F=F-Mg=M\frac{dV_{C.M.}}{dt}= 
-\mu (\dot{y}_2-\dot{y}_1)=\mu(V_1+V_2),   
\end{equation}
which is identical with (\ref{Delta_F}).

We hope the above discussion illustrates why the hourglass problem is an excellent pedagogical problem. It allows to demonstrate the main components of a good strategy for solving a complex physical problem: identify the main physical processes relevant to the problem, build a plausible model of the phenomenon under study, carefully use the laws of physics, common sense and ingenuity, make reasonable approximations, and finally try to get some of the key results in a different way to gain confidence that you are on the right track. 

However, as we show in this note, the pedagogical possibilities of the hourglass problem are not limited to what was described above. Namely, the hourglass problem can be used to demonstrate the use methods of variable mass systems. For this goal we use the modification of the problem described in Ref.~\cite{Becker2008}.

\section{Fraction of the sand in the free-fall}
\begin{figure}
    \centering
    \includegraphics[width=0.35\linewidth]{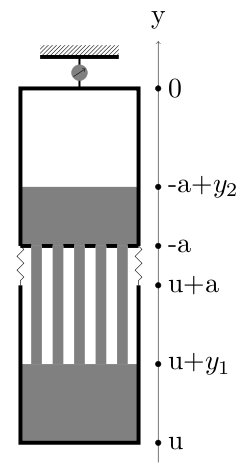}
    \caption{Schematic geometry of the hourglass used in the text. $a$ is the height of the cylinders. $y_1$ and $y_2$ are the heights of sand piles in the lower and upper compartments, respectively.}
    \label{fig1}
\end{figure}

We will assume that the hourglass consists of cylindrical upper and lower compartments of constant cross section, separated by a sieve. It was shown in Ref.~\cite{Becker2008} that for an hourglass to have a constant weight, it is necessary that the sand surface in the lower compartment be maintained at a constant height, and that this can be achieved if the lower container is suspended via springs, as shown in Fig.\ref{fig1}.

The geometric parameters are the same as in Ref.~\cite{Tuinstra_2010}. That is each cylinder has a diameter of 8~cm and a length of $a=20$~cm. The sieve consists of 230 holes with a diameter of 2.0~mm each. Coordinate conventions are similar as in Ref.~\cite{Becker2008}, and the general geometry is shown in Fig.\ref{fig1}.

Since it is not immediately clear that in the case of multi-holes, the mass of sand in free fall can be neglected, let us estimate the fraction of sand in the free-falling cascade, assuming a constant flow of sand $\mu$.  If $x$ is the downward distance from the level of sieve, the mass of the sand in air is given by
\begin{equation}
\Delta m=\int\limits_0^{\Delta l+(a-y_1)}\rho(x) A(x) dx,
\label{eq1}
\end{equation}
where $\Delta l=-2a-u$ is the elongation of the spring connecting upper and 
lower compartments. Note that the $y$-axis is directed upward and $u$ is the
coordinate of the bottom of the lower compartment, so $u<0$. As the flow 
$\mu$ is constant, we have at any cross-section
\begin{equation}
\mu=\rho(y) A(y) V(y),
\label{eq2}
\end{equation}
But at the cross-section which is situated at a distance $x$ below the 
sieve the free-fall velocity is
\begin{equation}
V=\sqrt{V_0^2+2gx},
\label{eq3}
\end{equation}
where $V_0$ is the velocity by which the sand leaves the sieve and enters the 
free-fall cascade. Therefore
\begin{equation} 
\Delta m=\int\limits_0^{-(a+u+y_1)}\frac{\mu\,dx}{\sqrt{V_0^2+2gx}}=
\frac{\mu V_0}{g}\left [\sqrt{1-\frac{2g}{V_0^2}(a+u+y_1)}-1\right ].
\label{eq4}
\end{equation}  
Note that we can rewrite this result in the following intuitively transparent 
way
\begin{equation}
\Delta m=\frac{\mu}{g}(V-V_0),
\label{eq5}
\end{equation}
where $V=\sqrt{V_0^2-2g(a+u+y_1)}$ is the velocity at the end of the 
free-fall cascade and $(V-V_0)/g$ is the free-fall time. 

Let us estimate $V_0$. If the total area of openings in the sieve is $A_0$,
then 
\begin{equation}
\mu=\rho A_0 V_0.
\label{eq6}
\end{equation}
Assuming that initially the sand completely fills the upper compartment, its 
total mass is $M=\rho A a$. But $\mu=M/T$, where $T$ is the time it takes to 
get the upper compartment empty. Therefore, (\ref{eq6}) gives
\begin{equation}
V_0=\frac{a}{T}\,\frac{A}{A_0}.
\label{eq7}
\end{equation}
For the hourglass used in Ref.~\cite{Tuinstra_2010}, $a/T\approx 0.84~\mathrm{cm/s}$ and
$$\frac{A}{A_0}=\frac{\pi\, 4^2}{230\, \pi\, 0.1^2}\approx 7.$$
Therefore, $V_0\approx 5.8 ~\mathrm{cm/s}$ and
$$\frac{2ga}{V_0^2}\approx 10^3.$$
Combined with (\ref{eq4}), this fact shows that a good approximation for
(\ref{eq4}) is (note that initially $u=-2a$ and $y_1=0$)
\begin{equation}
\Delta m\approx \mu\sqrt{-2\,\frac{a+u+y_1}{g}}
\label{eq8}
\end{equation}
and
\begin{equation}
\frac{\Delta m}{M}=\frac{1}{T}\sqrt{-2\,\frac{a+u+y_1}{g}}=\frac{a}{T}
\sqrt{\frac{-2(a+u+y_1)}{a^2g}}.
\label{eq9}
\end{equation}
For example, initially
\begin{equation}
\frac{\Delta m}{M}=\frac{a}{T}\,\sqrt{\frac{2a}{a^2g}}=\sqrt{2}\,\frac{a}{T}\,
\sqrt{\frac{1}{ag}}\approx 0.9\cdot 10^{-2}.
\label{eq10}
\end{equation}
Besides, from (\ref{eq4}) we get
\begin{equation}
\frac{d}{dt}\Delta m\sim (\dot{u}+\dot{y}_1).
\label{eq11}
\end{equation}
Below we will be interested in a situation when
$$\dot{y}_1=-\dot{u}=\frac{\mu}{\rho A},$$
and, therefore, (\ref{eq11}) shows that
\begin{equation}
\frac{d}{dt}\Delta m= 0.
\label{eq12}
\end{equation}
In conclusion, only a small fraction of the sand is in a free-fall during 
all the time and we can ignore, if needed, its mass compared to the masses 
in the hourglass compartments except, perhaps, very brief periods at the 
beginning and at the end of the flow. 

\section{The weight of the hourglass}
Hourglass, as a whole, is a constant-mass system. Therefore, we can write
\begin{equation}
\frac{dp_y}{dt}=F-Mg,
\label{eq13}
\end{equation}
where $F$ is the force from the balance spring and it just equals to the 
weight $P$ of the hourglass. If $x$ is the downward distance from the top 
level of the sand, then we have
\begin{equation}
p_y=-\int\limits_0^{-a+y_2-(u+y_1)}\rho(x)A(x)V(x)\,dx+(m_1+\tilde{m}_1)\dot{u},
\label{eq14}
\end{equation}
where $m_1$ is the mass of the sand in the lower compartment except the sand 
in a free fall and $\tilde{m}_1$ is the constant mass of the compartment 
itself. 

Maybe some clarification would be helpful here. The integration in equation (\ref{eq14}) runs over the sand in the upper container and the falling part of the sand. This part of the sand moves relative to the lower container, while the remaining part of the sand in the lower compartment of the (variable) mass $m_1$ moves along with its container.

Remembering (\ref{eq2})\footnote{Equation (\ref{eq2}) is a consequence of the mass conservation (or continuity) equation
$$\frac{\partial \rho}{\partial t}+\nabla \cdot(\rho \vec{V})=0$$
for a stationary flow with $\frac{\partial \rho}{\partial t}=0$. In this case, the surface integral $\oint\limits_\Sigma\rho \vec{V}\cdot d\vec{S}=0$. If $\Sigma$ is the lateral surface of the hourglass bounded from above by the cross-sectional area $A(y)$ of the hourglass at height $y$ and the bottom surface of the upper compartment, we get $\rho(y)A(y)V(y)-\rho A_0V_0=\rho(y)A(y)V(y)-\mu=0$. Therefore, equation (\ref{eq2}) is valid both in the sand fall zone and inside the upper compartment, although in the latter case, of course, $\rho(y)=\rho$.}, we get
\begin{equation}
p_y=-\mu (y_2-y_1-a-u)+(m_1+\tilde{m}_1)\dot{u}.
\label{eq15}
\end{equation}
Therefore,
\begin{equation}
\dot{p}_y=-\mu (\dot{y}_2-\dot{y}_1-\dot{u})+\dot{m}_1\dot{u}+
(m_1+\tilde{m}_1)\ddot{u}.
\label{eq16}
\end{equation} 
But $m_1+m_2+\Delta m=M$, where $m_2$ is the mass of the sand in the upper
compartment. In view of (\ref{eq12}) and because obviously $\dot{m}_2=-\mu$,
we have
\begin{equation}
\dot{m}_1=-\dot{m}_2=\mu.
\label{eq17}
\end{equation}
On the other hand,
\begin{equation}
y_1=\frac{m_1}{\rho A},\;\;\;y_2=\frac{m_2}{\rho A}.
\label{eq18}
\end{equation}
Therefore,
\begin{equation}
\dot{y}_1=\frac{\mu}{\rho A},\;\;\;\dot{y}_2=-\frac{\mu}{\rho A}.
\label{eq20}
\end{equation}
Substituting (\ref{eq17}) and (\ref{eq20}) in (\ref{eq16}), we get
\begin{equation}
\dot{p}_y=2\,\frac{\mu^2}{\rho A}+2\,\mu\dot{u}+(m_1+\tilde{m}_1)\ddot{u}.
\label{eq21}
\end{equation}
This is essentially the equation (6) from Ref.~\cite{Becker2008} and we see that
$\dot{p}_y=0$ (the hourglass has a constant weight $P=Mg$) if
\begin{equation}
\dot{u}=-\frac{\mu}{\rho A}.
\label{eq22}
\end{equation}
It is noteworthy \cite{Becker2008} that the condition $\dot{u}+\dot{y}_1=0$ means that the sand surface in the lower compartment is maintained at a constant height (see Fig.\ref{fig1}).

\section{Equation of motion for the lower compartment}
For variable mass systems neither $M\frac{d\vec{V}}{dt}=\vec{F}$ nor $\frac{d\vec{p}}{dt}=\vec{F}$ are the correct equations of motion in the general case, where $\vec{F}$ is an external force. Instead, the general situation is described by the equation \cite{Tiersten_1969}
\begin{equation}
\frac{d\vec{p}}{dt}=\vec{F}+\vec{\pi},    
\label{eq_pi}
\end{equation}
where $\vec{\pi}$ is the momentum flux into the system (how much momentum is brought into the system by new parts per unit time). Below, we use (\ref{eq_pi}) to re-derive and somewhat correct the equation of motion for the bottom part of the hourglass from Ref.~\cite{Becker2008}.

Let us consider a variable mass system which consists of the lower compartment with the sand in it plus the sand in the free fall. Its equation of motion has the form
\begin{equation}
\dot{p}_{1y}=\pi_y+k\Delta l-[m_1+\tilde{m}_1+\Delta m]g,
\label{eq23}
\end{equation}
where
\begin{equation}
\pi_y=-\mu V_0
\label{eq24}
\end{equation}
is the momentum flux in this system. But
\begin{equation} 
p_{1y}=-\int\limits_0^{\Delta l+(a-y_1)}\rho(x)AV(x)\,dx+[m_1+\tilde{m}_1]\dot{u}=
\mu(a+u+y_1)+[m_1+\tilde{m}_1]\dot{u}.
\label{eq25}
\end{equation}
Therefore, since $\Delta l=-2a-u$, equation (\ref{eq23}) gives
\begin{equation} 
[m_1+\tilde{m}_1]\ddot{u}+[\mu+\dot{m}_1]\dot{u}+ku= 
-[m_1+\tilde{m}_1+\Delta m]g-2ak-\mu(V_0+\dot{y}_1).\;\;\;
\label{eq26}
\end{equation}
If we neglect $\Delta m$ compared to $m_1$, then $\dot{m}_1=-\dot{m}_2=\mu$,
$\dot{y}_1=\mu/\rho A$ and (\ref{eq26}) reduces to
\begin{equation}
[m_1+\tilde{m}_1]\ddot{u}+2\mu\dot{u}+ku= 
-[m_1+\tilde{m}_1]g-2ak-\frac{\mu^2}{\rho A}-\mu V_0,
\label{eq27}
\end{equation}
which is identical to the equation (8) from Ref.~\cite{Becker2008} with $\eta=2\mu$ and $F_j=2ak+\mu^2/(\rho A)+\mu V_0$. As we can see, to obtain this equation, there is no need for the dissipation in the spring. The effective dissipation constant $\eta=2\mu$ arises dynamically partly from the term $\dot{p}_{1y}$, and partly from the momentum flux $\pi_y$.

\section{Another derivation of the equation of motion} 
The derivation of the equation of motion for the lower compartment becomes somewhat more complicated if we do not include sand in free fall in the variable mass system from the very beginning, taking the boundary of the system just above the bottom of the falling column of sand. In this case $p^\prime_{1y}=(m_1+\tilde{m}_1)\dot{u}$ and naively we can think that $\pi^\prime_y=-\mu V$. However the latter equation is not correct. The fact is that the surface in the lower compartment, where the sand ends its free-fall and stops, moves upward with the velocity $\dot{u}+\dot{y}_1$. Because of this motion, it takes time other than $dx/V$ for a portion of momentum $\Delta p_y=-\rho A V dx=-\mu dx$ to be absorbed by our variable mass system. Namely, the absorption time is
\begin{equation}
dt=\frac{dx}{V+\dot{u}+\dot{y}_1}
\label{eq28}
\end{equation}
and, therefore,
\begin{equation}
\pi^\prime_y=-\mu (v+\dot{u}+\dot{y}_1).
\label{eq29}
\end{equation}
Hence the equation of motion
\begin{equation}
\dot{p}^\prime_{1y}=\pi^\prime_y+k\Delta l-[m_1+\tilde{m}_1]g
\label{eq30}
\end{equation}
takes the form
\begin{equation} 
[m_1+\tilde{m}_1]\ddot{u}+\dot{m}_1\dot{u}= 
-\mu(V+\dot{u}+\dot{y}_1)+k\Delta l-[m_1+\tilde{m}_1]g.
\label{eq31}
\end{equation}
But, from (\ref{eq5}), $\mu V=\mu V_0+\Delta m\, g$ and we see that 
(\ref{eq31}) is identical to the equation (\ref{eq26}).

\section{The case of the zero momentum flux}
What happens if the boundary of the variable mass system is taken just below the bottom of the falling column of sand? In this case, the new parts of the system are already stationary and, therefore, the momentum flux is zero. However, instead of the momentum flux, in this case there will appear a downward impact force $F$ resulting from the impact of freely falling particles on the boundary surface of the variable mass system. This is exactly the situation considered in Ref.~\cite{Becker2008}. The equation of motion is
\begin{equation}
\dot{p}^\prime_{1y}=k\Delta l-[m_1+\tilde{m}_1]g-F,
\label{eqa1}
\end{equation}
where, as in the previous chapter, $p^\prime_{1y}=(m_1+\tilde{m}_1)\dot{u}$. From what was said in the previous chapter, it is easy to calculate the impact force $F$. Indeed, a small part of falling sand with momentum $\Delta p_y=-\rho A V dx=-\mu dx$ is stopped in time $\Delta t=dx/(V+\dot{u}+\dot{y}_1)$. Hence $F=(0-\Delta p_y)/\Delta t=\mu(V+\dot{u}+\dot{y}_1)=-\pi^\prime_y$ and equation (\ref{eqa1}) becomes identical to the equation (\ref{eq30}).

Note that
$$F=\mu(V+\dot{u}+\dot{y}_1)= \left (\mu V_0+\frac{\mu^2}{\rho A}+\Delta m g\right )+\mu\dot{u}$$
contains not only the constant part in parentheses ($\Delta m$ is only asymptotically constant. In any case, it can be neglected in the equation of motion due to its smallness compared to other masses, as mentioned above), but also $\mu\dot{u}$ term. This last term was overlooked in Ref.~\cite{Becker2008}, just as its twin term that originates from $\dot{p}^\prime_{1y}$. The latter circumstance is due to the fact that Ref.~\cite{Becker2008} in (\ref{eqa1}) instead of $\dot{p}^\prime_{1y}$ uses the Newtonian $(m_1+\tilde{m}_1 )\ddot {u}$.

It was shown in Ref.~\cite{Becker2008} that if $k=\rho A g$, then the condition (\ref{eq22}) is satisfied asymptotically and the hourglass has constant weight, provided $\eta/\mu>1/2$. For an effective dissipation constant $\eta=2\mu$, this stability condition is satisfied automatically and there is no need for dissipation in the spring, unless we want to further increase the stability.

\section{Otto-McDonald leaky tank car}
The general theory of systems with variable mass is rather complicated \cite{Eke_2002,Casetta2016}. However, the concept of the momentum flux and the equation (\ref{eq_pi}) make it much easier to obtain the equation of motion for various systems with variable mass. For example, let us demonstrate this by obtaining the equation of motion for a leaky tank car \cite{McDonald_1991,McDonald_2018}.

\begin{figure}
    \centering
    \includegraphics[width=0.8\linewidth]{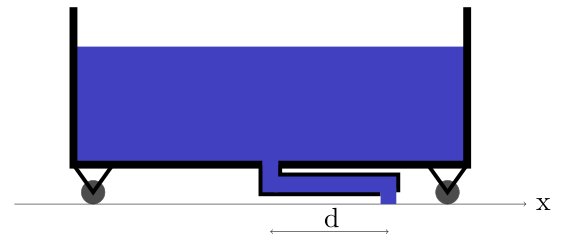}
    \caption{Otto-McDonald model of internal motion of water into a leaky tank car. }
    \label{fig2}
\end{figure}

There are two principal aspects in the problem of a leaky tank car. Firstly, the tank car is a variable mass system with a non-zero momentum flux, and secondly, there is an internal horizontal motion of water\footnote{The water can be replaced by a fluidized granule.} in the tank. The Otto-McDonald model shown in Fig.\ref{fig2} makes this internal motion explicit. 

Namely, let us imagine that an off-center drain is connected by a pipe of length $D$ and cross-sectional area $S$ with a hole in the center of the bottom of the tank. The water in the pipe moves relative to the tank with the velocity $u$ and hence with the velocity $\dot{x}+u$ relative to the fixed inertial reference frame, in which the coordinate of the center of mass of the tank car is equal to $x$.

 For an infinitesimally small time interval $dt$, the amount of water $dm=\rho Su dt$ leaves the tank with the same horizontal speed $\dot{x}$ (with respect to a fixed inertial frame of reference) that the tank has at that moment. Therefore, the amount of horizontal momentum leaving the reservoir in the time $dt$ is equal to $dp=dm \dot{x}$. Since the momentum flux in (\ref{eq_pi}) is defined as the amount of momentum brought by the new parts of the system, we conclude that in our case 
 \begin{equation}
 \pi_x=-\frac{dp}{dt}=-\rho S u\dot{ x}.
 \label{pi_x}
 \end{equation}
 
On the other hand, the horizontal momentum of the tank car is
\begin{equation}
p_x=(M-\rho SD)\dot{x}+\rho SD(\dot{x}+u)=M\dot{x}+\rho SDu,
\label{p_x}
\end{equation}
where $M$ is the total mass of the tank car (including water).

External horizontal force on the tank car equals zero. Therefore, (\ref{eq_pi}) becomes $\frac{dp_x}{dt}=\pi_x$, and substituting (\ref{pi_x}) and (\ref{p_x}) into this equation, we finally obtain
\begin{equation}
M\ddot{x}=-\rho SD\dot{u}=D\ddot{M},
\label{eq_tank}    
\end{equation}
where we have taken into account that $\dot{M}=-\rho Su$. The equation (\ref{eq_tank}), as the equation of motion of a leaky tank car, was first obtained and investigated in Ref.~\cite{McDonald_1991}.

\section{Concluding remarks}
Variable mass systems and systems of variable composition play an important role in modern technology \cite{Cveticanin_1998} and astrophysics \cite{Hadjidemetriou_1967}. Despite the fact that the dynamics of variable mass systems has a venerable history and has been an active area of research for many years \cite{Eke_2002}, this branch of mechanics is usually given little attention in textbooks, and one can still find in the literature incorrect applications of Newton's second law in this context \cite{Plastino_1992}.

Textbook treatments of systems with variable mass, as a rule, are limited to the movement of a rocket and a discussion of the Meshchersky equation and reactive forces. Unfortunately, this does not give the student enough intuition to solve such interesting variable mass problems as the motion of a leaky tank car \cite{McDonald_1991,McDonald_2018,Esposito2022} or falling chains \cite{Denny_2020,Silagadze_2010}.

In conclusion, the concept of momentum flux is a useful concept in the study of variable mass systems and it is desirable that students be familiar with this concept. The hourglass problem, as we have tried to show in this note, gives students one more opportunity to practice understanding this useful concept.

\section*{Acknowledgments}
Johann Otto drew our attention to this interesting variable mass problem many years ago. We thank Thorsten P\"{o}schel and anonymous referee for helpful comments.
The work is supported by the Ministry of Education and Science of the Russian Federation.

\bibliography{hourglass}

\end{document}